\documentstyle[preprint,prl,aps]{revtex}
\begin{document}
\tighten
\title{Probing proton halos through pion photoproduction}
\author{S.~Karataglidis\cite{newad}}
\address{National Superconducting Cyclotron Laboratory\\ 
Michigan State University, East Lansing, Michigan, 48824-1321}
\author{C.~Bennhold}
\address{Center of Nuclear Studies, Department of Physics\\
The George Washington University\\
Washington, D. C. 20052}
\date{\today}
\maketitle
\begin{abstract}
Charged pion photoproduction is proposed as a new method to study
halo nuclei. In particular, it is demonstrated that the reaction
$^{17}$O($\gamma,\pi^-$)$^{17}$F$(\frac{1}{2}^+$, 0.495 MeV) 
is well-suited to investigate the well-known proton halo in $^{17}$F
which is not amenable to radioactive beam experiments.  The cross 
sections are shown to be very sensitive to the halo structure of the
loosely bound proton.
\end{abstract}

Much information on the momentum distributions of halo nucleons in
halo nuclei has been gained from heavy ion collisions. One of the most
striking experimental feature of reactions in which a halo system is
broken up is the extremely narrow momentum distribution of the
fragments, reflecting the spatially extended nature of the halo wave
functions with r.m.s radii of $5-7$~fm.  Among the best-studied cases of
neutron-rich nuclei are $^{11}$Be as a single-neutron halo and
$^{11}$Li as a loosely bound two-neutron halo. The analogues of these
systems near the proton drip line are all unbound. Proton halos are
hampered by the additional Coulomb barrier outside the nuclear surface
that tends to cut off the tail of the proton wave function.

The best candidates for proton halos are those formed by valence
protons in $s$ states \cite{Br96}.  Such an example with a clear
signature for a proton halo is the first excited $\frac{1}{2}^+$ state
of $^{17}$F; the separation energy of the valence proton is only
105~keV \cite{Ti93}. This state plays an important role in stellar
nucleosynthesis since radiative proton capture on $^{16}$O is
responsible for the breakout from the carbon-nitrogen-oxygen (CNO)
cycle \cite{Ro73}.  This transition proceeds from a continuum
$p$-state to this excited state and is strongly enhanced by the tail
of the valence proton of the bound state. Such a tail is the typical
signature of a halo structure. The halo nature of the proton has also
been inferred from the large Thomas-Ehrman Coulomb shift relative to
the $\frac{5}{2}^+, T=\frac{1}{2}$ ($0d_{\frac{5}{2}}$) ground state
\cite{No69,Sh85} and from the large 1/2$^+$ to 5/2$^+$ B($E2$) value
\cite{Br77}.  Furthermore, in a study \cite{Bo93} of the first
forbidden $\beta$-decay of $^{17}$Ne to the excited state of $^{17}$F
the decay rate was found to be almost a factor two above the rate
obtained from the well-known mirror decay of $^{17}$N to the first
excited state of $^{17}$O.  While this largest mirror asymmetry for
first-forbidden decays between bound states recorded to date was
considered to be clear evidence of an extended proton structure in
$^{17}$F, Millener \cite{Mi97} argued that these differences can be
explained in terms of charge-dependent effects. These charge-dependent
effects themselves also arise from the modifications of the radial
wave functions of the valence nucleons due to the loose binding of the
$s$-state valence proton \cite{No69}.

The reason for the difficulties in the interpretation of this (excited
state) single proton halo is that its reaction cross section and
momentum distribution cannot be studied in radioactive beam
experiments. Herein, we propose to use the reaction
$^{17}$O($\gamma,\pi^-$)$^{17}$F$(\frac{1}{2}^+$, 0.495 MeV) to
extract the proton halo structure of this state. In principle, halo
states may be formed by single charge exchange reactions on stable
nuclei that fall into two broad classes: those mediated by the strong
interaction [($p,n$), ($n,p$), or $(\pi^{\pm},\pi^0$)] and those
mediated by the electromagnetic interaction
[($\gamma,\pi^{\pm}$)]. The latter are favored for the study of halo
states as they are soft electromagnetic reactions, and can largely
preserve the momenta of the valence nucleons. The purpose of this
investigation is not only to demonstrate the feasibility of pion
photoproduction experiments leading to halo nuclear states, but also
to show that such experiments will indeed be sensitive to the details
of the wave functions of the valence nucleons. We stress that this
investigation is exploratory and meant to stimulate theoretical
discussion and experimental efforts. Ideally, one would also
require complementary data from nucleon or pion charge exchange
reactions leading to the same halo states. Together with the photopion
data, they would provide the most sensitive tests available for the
halo wave functions.

In the past 20 years, pion photoproduction from light nuclei has
developed into an important tool in addition to electron and proton
scattering to examine the nuclear structure of light
nuclei\cite{Na91}.  The four ingredients in the theoretical
description of the ($\gamma,\pi^{\pm}$) reaction are: 1) the nuclear
structure transition density matrix elements, 2) the amplitude for the
elementary pion photoproduction process off a free nucleon, 3) the
single-particle shell-model wave functions, and, 4) the final state
interaction of the outgoing pion with the residual nucleus. Using the
notation of Ref. \cite{Ti84} the many-body matrix elements may be
written as
\begin{equation}
\left\langle J_f M_f T_f P_f; \pi \left| T \right| J_i M_i T_i P_i ;
\gamma \right\rangle 
= \sum_{\alpha,\alpha^{\prime}} \left\langle J_f M_f T_f P_f \left|
a^{\dagger}_{\alpha^{\prime}} a_{\alpha} \right| J_i M_i T_i P_i
\right\rangle \left\langle \alpha^{\prime} ; \pi \left| t \right|
\alpha; \gamma \right\rangle \; ,
\end{equation}
where $\alpha$ denotes $\{n,l,j,m,\rho\}$ and the indices $i$ and $f$
refer to the initial and final nuclear states.  The nuclear structure
matrix element is expanded in terms of the one-body transition density
matrix elements (OBDME), reduced in both spin and isospin, i.e.
\begin{equation}
\Psi_{JT}(\alpha^{\prime},\alpha) = \frac{1}{\sqrt{(2J+1)(2T+1)}} \left\langle
J_f T_f \left|\left|\left| \left[ a^{\dagger}_{\alpha^{\prime}} \times
\tilde{a}_{\alpha} \right]^{JT} \right|\right|\right| J_i T_i
\right\rangle.
\end{equation}
\par
The one-body transition operator, $t$, is given by
\begin{eqnarray}
t & = & \left( L + i \vec{\sigma} \cdot \vec{K} \right)
\frac{\tau_{-\beta}}{\sqrt{2}} \\
& = & \sum_{S,M_S} i^S (-1)^{M_S} \sigma^S_{-M_S} K^S_{M_S}
\frac{\tau_{-\beta}}{\sqrt{2}},
\end{eqnarray}
where $L$ and $\vec{K}$ are the spin 0 and spin 1 transition operators
\cite{Ti84}, respectively.  As the elementary pion production
amplitude we choose the simple operator of Blomqvist and Laget (BL)
\cite{Bl77,La87} which has been designed specifically with nuclear
applications in mind. While more sophisticated approaches have been
developed in recent years \cite{Su96} which give a good description of
the $(\gamma,\pi)$ multipoles they are less suitable for nuclear
applications since their off-shell extrapolation is not
straightforward. The BL amplitude incorporates the essential features
and gives an adequate description of the N$(\gamma,\pi)$N cross
section data. The BL operator consists of tree-level Born terms that
include the model-independent Kroll-Ruderman (KR) term, as well as the
contribution from the $\Delta(1232)$. Future studies should use the
more advanced production operators that adequately include unitarity
dynamically.

The matrix element of this operator between bound nuclear states is
\begin{equation}
\left\langle \alpha^{\prime} ; \pi \left| t \right| \alpha; \gamma
\right\rangle = \int \; d^3p \, d^3q^{\prime} \;
\Psi^{\ast}_{\alpha^{\prime}}(\vec{p^{\prime}})
\varphi^{(-)\ast}_{\pi}(\vec{q^{\prime}},\vec{q})
t_{\gamma,\pi}(\vec{p},\vec{p^{\prime}},\vec{k},\vec{q^{\prime}})
\Psi_{\alpha}(\vec{p})
\end{equation}
where $\vec{p}$ is the momentum of the initial nucleon and
$\vec{p^{\prime}} = \vec{p} + \vec{k} - \vec{q^{\prime}}$ from momentum
conservation. The outgoing (distorted) pion wave function, 
obtained as a solution of the Schr\"{o}dinger equation with the
optical potential of Stricker, McManus and Carr \cite{St79}, is
denoted by $\varphi^{(-)\ast}_{\pi}(\vec{q^{\prime}},\vec{q})$.

The matrix element may be expressed as \cite{Ti84}
\begin{eqnarray}
\left\langle \alpha^{\prime}; \pi \left| t \right| \alpha; \gamma
\right\rangle & = & \sum_{LSJM} i^S
(-1)^{j-m+\frac{1}{2}-\rho^{\prime}+l^{\prime}+s} \sqrt{6}
\sqrt{(2j+1)(2j^{\prime}+1)(2L+1)(2S+1)} \nonumber \\
& \times &
\left( \begin{array}{ccc}
	\frac{1}{2} & \frac{1}{2} & 1 \\
	-\rho^{\prime} & \rho & -\beta
	\end{array} \right)
\left( \begin{array}{ccc}
	j' & j & J \\
	-m' & m & M
	\end{array} \right)
\left\{ \begin{array}{ccc}
         l' & \frac{1}{2} & j' \\
         l  & \frac{1}{2} & j  \\
         L  & S           & J
	\end{array} \right\}
I^{(a^{\prime}a)LSJ}_{-M} \;.
\end{eqnarray}
In $LS$-coupling the radial integral, $I^{(a^{\prime}a)LSJ}_M$,
therein is given by
\begin{equation}
I^{(a^{\prime}a)LSJ}_M = \int \; d^3p \, d^3q^{\prime} \;
\varphi^{\ast}_{n^{\prime}l^{\prime}j^{\prime}}(p^{\prime})
\varphi_{nlj}(p) \varphi_{\pi}^{(+)}(\vec{q^{\prime}},\vec{q}) \left[
\left[ Y_{l^{\prime}m_l^{\prime}}(\hat{p}^{\prime}) \times
Y_{lm_l}(\hat{p}) \right]^L \times K^S \right]^{JM}.
\end{equation}
This integral is dependent on the single particle wave functions of
the outgoing pion, and of the initial and final state nucleons
involved in the transition.  It is through this integral that the halo
wave function enters into the calculation.  Hence the pion
photoproduction reaction probes the whole wave function of the halo
nucleon. This is in contrast to measurements of the momentum
distributions of the halo nucleons in heavy ion collisions, where the
part of the wave function inside the core is hidden \cite{Ha96}. 

To illustrate the sensitivity of the cross section to the halo nature
of the valence proton, we contrast the difference between the results
obtained using the harmonic oscillator (HO) and Woods-Saxon (WS) single
particle wave functions.  As the charge radius for $^{17}$O is the
same as that for $^{16}$O \cite{Ki78}, the HO parameter for the
present calculations was 1.7~fm, as obtained from an analysis of the
elastic electron scattering form factor for $^{16}$O \cite{Ka96}. The
WS wave function for the $1s_{\frac{1}{2}}$ proton was obtained by
solving the Schr\"odinger equation with a Woods-Saxon potential that
reproduces the single particle binding energy of 105~keV, the
separation energy of that proton \cite{Ti93}.  The difference between
the single particle densities obtained using the two wave functions is
shown in Fig.~\ref{wavefn}.  In contrast to the HO density the tail of
the WS wave function exhibits the dramatic enhancement of the tail,
characteristic of halo particles, and is similar to single-neutron
halos such as the $1s_{\frac{1}{2}}$ state of $^{11}$Be.  The
r.m.s. radius of our WS wave function is 5.1~fm.  The enhancement at
large $r$ requires a reduction of the strength at short distances to
preserve normalization.  Note that the node occurs at around 2~fm for
both HO and WS densities.

The $^{17}$O and $^{17}$F mirror nuclei are described by pure single
particle states outside a closed $^{16}$O core in the $0\hbar\omega$
shell model. This is a simple model: the transverse magnetic
form factor for the elastic scattering of electrons from $^{17}$O is
suppressed compared to the predictions of the single particle model
\cite{Zh92}. Yet the magnetic moment is $-1.894\; \mu_N$
\cite{Ti93}, which is very close to that of the neutron
($-1.913\;\mu_N$). Our simple model may then be valid as a first
approximation. The 0.495~MeV excited state therefore has a much simpler
proton halo configuration compared to $^8$B. As such, the
$^{17}$O($\gamma,\pi^-$)$^{17}$F reaction is effected by a pure
$0d_{\frac{5}{2}} \rightarrow 0d_{\frac{5}{2}}$ (ground state) or
$0d_{\frac{5}{2}} \rightarrow 1s_{\frac{1}{2}}$ (excited state) single
particle transition. In each case the isovector OBDME, in $jj'$--coupled form,
are unity for all given multipolarities.

The results of our calculations of the
$^{17}$O($\gamma,\pi^-$)$^{17}$F differential cross sections, for
$E_{\gamma} = 200$~MeV, are displayed in Fig.~\ref{o17g}(a). The result
for the ground state cross section obtained using the HO wave function
is displayed by the dashed line while that obtained using the WS wave
function is displayed by the dotted line. The binding energy for the
WS wave function in this case was set to the separation energy,
600~keV \cite{Ti93}, of the proton in the ground state.  The result of
the ground state cross section at forward angles is pronounced, around
3~$\mu$b/sr, decreasing to $\sim 1$~$\mu$b/sr at backward angles. This
enhancement at forward angles may be explained by the exact overlap of
the $0d_{\frac{5}{2}}$ neutron and proton single particle wave
functions. There is only a negligible difference between the HO and WS
results indicating that the ground state is not a halo, as
expected. The WS wave function does not have a pronounced tail in this
case.

The results of the cross section for the transition to the excited
state obtained using the HO and WS wave functions are displayed in
Fig.~\ref{o17g}(a) by the solid and dot-dashed lines respectively. The
cross section is suppressed at forward angles by the orthogonality of
the $0d_{\frac{5}{2}}$ neutron in $^{17}$O with the $1s_{\frac{1}{2}}$
proton in $^{17}$F$^*$. However, at backward angles the cross section
rises to around 1~$\mu$b/sr. Using the halo WS wave function instead of
the standard HO density reduces the excited state cross section at
large angles by a factor of two. That dramatic reduction at large momentum
transfer can be traced to the reduction of the $1s_{\frac{1}{2}}$
radial wave function at short distances, as illustrated in
Fig.~\ref{wavefn}.

In order to cleanly extract the proton halo wave function one requires
as little uncertainties from the pion photoproduction process as
possible. Hence, we compare plane wave impulse approximation (PWIA)
and distorted wave impulse approximation (DWIA) results for the excited
state cross section, as obtained using the HO wave function, in
Fig.~\ref{o17g}(b). The pion final state interaction at these energies
affects the cross section by at most 10\%; furthermore, the
uncertainty due to pion distortion would be only a fraction of
this. Thus the size of this effect is much smaller than the
sensitivity to the halo wave function.

Much of the excited state cross section comes from the KR term. The
other Born terms provide most of the remaining cross section, with at
most 10\% coming from the $\Delta$(1232).  It has been the latter term
which has lead to strong model dependencies in reactions like
$^{14}$N($\gamma,\pi^+$)$^{14}$C(g.s.) at higher photon energies
\cite{Mu86,Ti88}. With a photon energy of 200 MeV one remains well
below the kinematic region in which the $\Delta$ resonance is excited.
The dominance of the KR term can be traced to that of the $M3$
multipole in the transition, providing about 70\% of the cross section
(as a $\frac{5}{2}^+ \rightarrow \frac{1}{2}^+$ transition, the $E2$ and
$M3$ nuclear multipoles can contribute), and it is this multipole which
is known to be almost totally dominated by the KR term
\cite{Ti84,Be90}. Therefore, we may conclude that uncertainties in the
elementary amplitude are minimal and will not obscure the extraction
of the halo state. Clearly, any remaining small uncertainties from nuclear
structure or the elementary operator would affect both calculations
equally and therefore would not alter our findings.

The dramatic sensitivity to the halo structure in $^{17}$F$^*$ and the
size of the differential cross section makes the reaction
$^{17}$O($\gamma,\pi^-$)$^{17}$F reaction an excellent candidate with
which to study the halo experimentally at low-energy monoenergetic
photon facilities, such as that at Duke, or using the $(e,e'\pi^+)$
reaction as is possible at Mainz.  The comparable magnitude of the
ground and excited state cross sections for backward angles will help
ease the separation between the two transitions with the appropriate
high resolution detectors.  The simplifications employed in this
investigation may influence the overall magnitude of the cross section
and further work is required in order to establish that
magnitude. However, as those approximations were applied equally to
both the halo and non-halo calculations, the effect of the halo is
real. We point out that both the sensitivity and the cross sections
are much larger in $^{17}$O($\gamma,\pi^-$)$^{17}$F than in comparable
processes that lead to neutron-halo nuclei, such as
$^{11}$B($\gamma,\pi^+$)$^{11}$Be or $^{15}$N($\gamma,\pi^+$)$^{15}$C
\cite{Lu97}.

In conclusion, we have demonstrated that the reaction
$^{17}$O($\gamma,\pi^-$)$^{17}$F(0.495 MeV)
displays a very clear signature of the proton halo in
$^{17}$F which cannot be studied in radioactive beam experiments.
Thus, the charged pion photoproduction reaction at low energy provides an
alternative means by which to study the momentum distributions of the
halo nucleons in halo nuclei. In particular, these reactions can probe
the effect of the wave function of the halo particle inside the core,
a region not accessible to heavy ion collisions. 
The study of halo nuclei through pion photoproduction
therefore has significant potential to expose unique signatures
of these exotic structures.

We thank B.A. Brown for helpful discussions. The work of CB is
supported by DOE grant DE-FG02-95-ER40907, while the work of SK is
supported by NSF Grants Nos. PHY-9403666 and PHY-9605207.

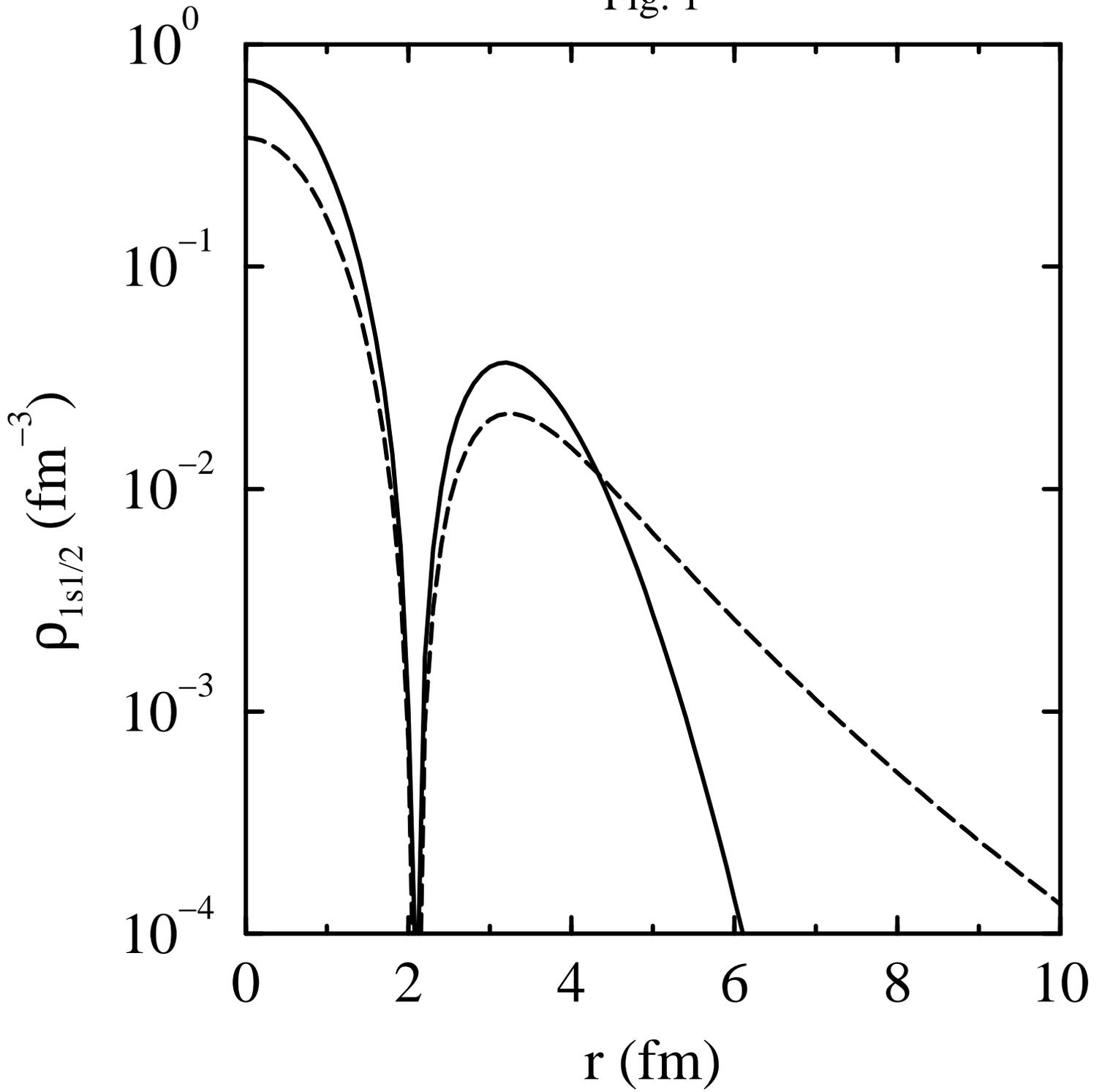
\begin{figure}
\caption[]{Radial wave functions, $R_{nlj}(r)$, for the
$1s_{\frac{1}{2}}$ orbit in $^{17}$O. The HO ($b=1.7$~fm) and WS ($E_B
= 105$~keV) wave functions are displayed by the solid and dashed lines
respectively.}
\label{wavefn}
\end{figure}
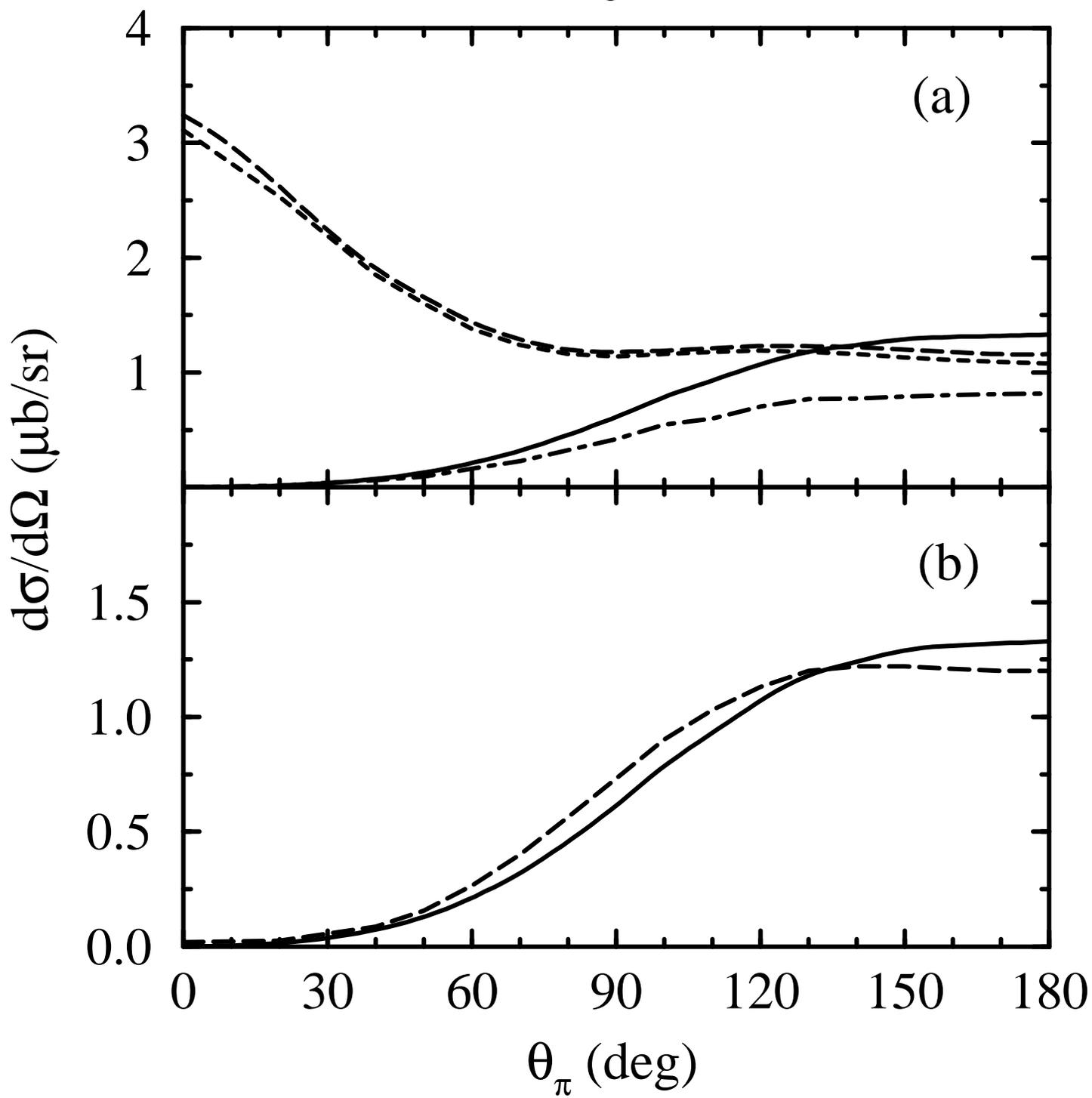
\begin{figure}
\caption[]{Differential cross section for
$^{17}$O($\gamma,\pi^-$)$^{17}$F at $E_{\gamma} = 200$~MeV. The cross
sections leading to the ground and first excited state of $^{17}$F are
displayed in (a) wherein the HO calculation for the excited state is
displayed by the solid line, while the WS result is displayed by the
dot-dashed line. The ground state cross section obtained using the HO
and WS wave functions are displayed by the dashed and dotted lines
respectively. The PWIA and DWIA results for the excited state cross
section obtained using the HO wave functions are displayed in (b) by
the solid and dashed lines respectively.}
\label{o17g}
\end{figure}
\end{document}